# Cluster size distributions in gas jets for different nozzle geometries

Márk Aladi,[a] Róbert Bolla,[a] Daniel E. Cardenas,[b] László Veisz,[b, c] and István B. Földes[a]

[a] *Wigner Research Centre for Physics, Hungarian Academy of Sciences,*
*1121 Budapest, Konkoly-Thege Miklós út 29-33., Hungary*

[b] *Max-Planck-Institut für Quantenoptik,*
*D-85748 Garching, Germany*

[c] Department of Physics, *Umeå University,*
*SE-901 87 Umeå, Sweden*
*E-mail*: `istvan.foldes@wigner.mta.hu`

Cluster size distributions were investigated in case of different nozzle geometries in argon and xenon using Rayleigh scattering diagnostics. Different nozzle geometries result in different behaviour, therefore both spatial- and temporal cluster size distributions were studied to obtain a well-characterized cluster target. It is shown that the generally used Hagena scaling can result in a significant deviation from the observed data and the behaviour cannot be described by a single material condensation parameter. The results along with the nanoplasma model applied to the data of previous high harmonic generation experiments allow the independent measurement of cluster size and cluster density.

## 1. Introduction

The use of pulsed gas jets gains more and more applications in different fields of the interactions of intense femtosecond lasers with matter. Noble gas jets are widely used for electron acceleration experiments, either in the "dream beam experiments" [1] or recently for shock front injection of electrons for further acceleration [2]. Efficient electron acceleration however requires relatively low densities in atomic gases.

Different is the case of generation of high harmonics (HHG) and thus attosecond XUV pulses [3]. The cluster-containing jets can be efficient targets for HHG [4, 5] or for the investigation of laser-driven cluster explosions [6]. After the initial great interest on HHG from clusters the interest declined mainly because the intensity limit on the generating pulse, and consequently the obtainable XUV energy can even be lower than for neutral gases [5, 7]. Another problem is the difficult reproducibility of the results which depends very strongly on the gas jet parameters, i.e. not only on the atomic density but also on the sizes of clusters and the density of the clusters, too. Besides requiring a flat-top (or at least Gaussian) density profile these parameters have to be controlled as well. Most of the works however use a simple scaling law of Hagena [8] for the cluster size. Rayleigh scattering is an excellent method for experimental characterization of the gas jets, and even recently detailed investigations were carried out for Ar clusters using this diagnostic method [9-11].

In recent experiments [7, 12] the presence of nanoplasmas in the clusters was demonstrated and based on a simple theoretical model of Tisch [13] even the density and the size of the clusters could be estimated from a spectral redshift of the harmonics with increasing backing pressure. In order to confirm the validity of the method and improve the accuracy it is necessary to compare the data with Rayleigh scattering. Additionally it must be noted that most of the previous diagnostics for the determination of cluster size were carried out in Ar only, although Xe can build larger clusters which is more advantageous for the nanoplasma phenomenon.

In the present paper different shapes and types of nozzles are characterized determining cluster sizes using Rayleigh scattering diagnostics. Besides giving a clear picture for the spatial distribution of the cluster formation in the jet the results are compared with the scaling laws showing that the generally applied analytic scaling laws can be strongly misleading.

## 2. Cluster size estimations for different nozzle geometries

In the adiabatic expansion of gas jets into vacuum Van der Waals bound atomic clusters are formed. The geometry of the nozzles has an important role for the efficiency of cluster formation. The average cluster size $N_c$ (number of the atoms / cluster) is generally estimated by the semi-empirical Hagena's parameter $\Gamma^* = k \cdot p_0 \cdot d_{eq}^{0.85} \cdot T_0^{-2.29}$ [8], where $k$ is the condensation parameter depending on the gas, $p_0$ the gas jet backing pressure in mbar, $T_0$ the gas temperature before expansion in Kelvin, and $d_{eq}$ the parameter for the nozzle geometry in µm. We can distinguish between sonic or supersonic nozzles. The Mach number increases in the flowing cold gas, and the increase is mainly caused by the dropping temperature, not so much the flow velocity [14]. For cylindrical sonic nozzles $d_{eq} = d_s$, where $d_s$ is the orifice diameter. For nozzles with conical geometry $d_{eq} = 0.74 d/\tan \alpha$ [8], where $d$ is the throat diameter and $\alpha$ is the half opening angle of the nozzle.

Hagena et al. introduced a scaling law for $\Gamma^* > 1000$ both for noble- and molecular gases. The size of the clusters scales with the backing pressure as $N_c \sim p_0^{2.35}$. In case of $10^4 < \Gamma^* < 10^6$ i.e. for higher backing pressure Dorchies et al. [9] proposed a corrected Hagena's law with a weaker scaling $N_c \sim p_0^{1.8}$ for conical nozzle. In order to obtain supersonic gas streams it is advantageous to use de Laval nozzles. Conical nozzles were investigated in several papers [9-11], but the de Laval geometries received less attention. On the one hand this is due to the difficulty of fabrication which requires an electro-erosion technique, on the other hand the experiments with de Laval nozzles haven't provided definite scaling data [9]. In the present paper we compare data obtained with different nozzles, including de Laval nozzles with conical divergent section.

## 3. Experimental

### 3.1 Nozzles

Throughout the experiments a commercial electromagnetic pulse valve (Parker series 9) was used with four different nozzles of the following parameters, see Figure 1(a):
(1) The cylindrical nozzle is a commercial nozzle with a straight, cylindrical orifice (Parker series 9), $d_s = 1000$ µm.
(2) The supplemental converging nozzle is a commercial nozzle (1) with a cylindrical supplement and a converging exit $d_s = 650$ µm.
(3) The de Laval nozzle has a conical divergent part with an exit diameter of 1 mm, $\alpha = 7°, d = 333$ µm.
(4) The de Laval micronozzle has a conical divergent part with an exit diameter of 0.15 mm, $\alpha = 7°, d = 50$ µm.

The total gas density from nozzles similar to (1) and (2) were earlier characterized by x-ray pulsed radiography technique in [15]. The density of gas jets from supersonic de Laval nozzles (3) and (4) were characterized with Mach-Zehnder interferometry in [14]. In the comparative calculations the cylindrical nozzles (1) and (2) were handled as sonic nozzles, whereas the de Laval nozzles (3) and (4) as conical nozzles. The solenoid valve was placed inside a vacuum chamber. The nozzles were directly attached to the valve for the sake of undisturbed gas flow except nozzle (2), for which the supplemental nozzle was fitted onto the commercial one. The investigation of the cluster formation with this fitting solution is reasonable, because it is a frequently used simple method to adopt different nozzle geometries [10, 15].

Our measurements were carried out at room temperature ($T_0 = 295$ K). The backing pressure was varied between 2 and 20 bar both with Ar ($k \approx 1650$) and with Xe ($k \approx 5500$).

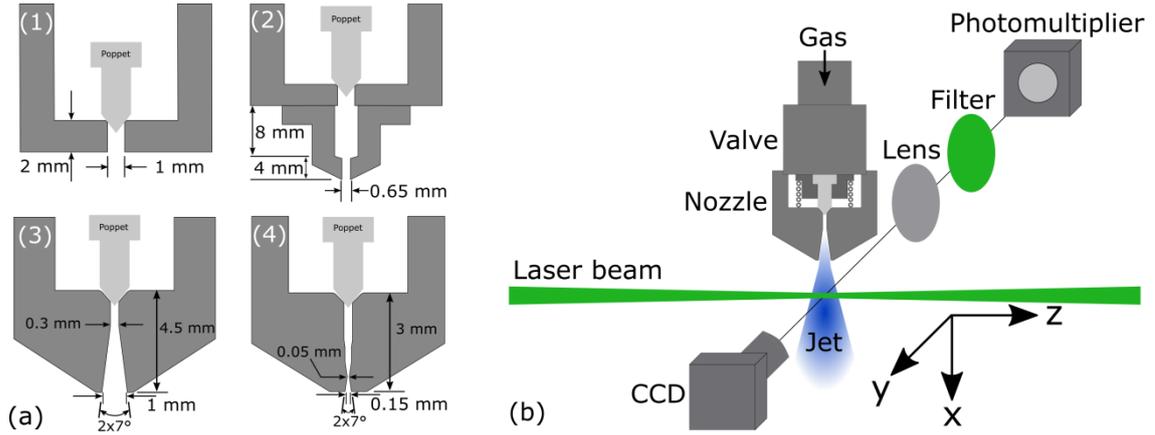

**FIG. 1.** (a) Sketch of the nozzles used in the experiments. (b) Schematic diagram of the experimental setup.

### 3.2 Rayleigh scattering

Laser Rayleigh scattering is an effective, relatively simple diagnostic tool for the investigation of gas properties and cluster formation in gas jets, too. From Rayleigh signals the absolute cluster size and density cannot be determined separately, nevertheless, with some assumptions a good estimation can be applied and also estimation of the absolute size is possible with simple calculations used by many groups [11]. A widely accepted assumption that for Rayleigh scattering the minimum detectable cluster size is approximately 100 atoms/cluster when the signal-to-noise ratio is about two. The signal was first reduced with the background level from stray light and dark current of the detector which can be observed in vacuum. Valve operation with very low (~0.1 bar) backing pressure (in which case no clusters are expected to be present) served as the noise level for determining the S/N ratio. A cluster radius can thus be estimated assuming Van der Waals bond for a given crystalline noble gas.

The Rayleigh scattering signal $S$ scales with the pressure as $S \sim p_0^\beta$. It can be shown [10, 11] - for spherical clusters and supposing that all the atoms are condensed into clusters - that the average cluster size $N_c \sim p_0^{\beta-1}$. The comparison of nozzles with different jet sizes require the knowledge of the interaction length, so $N_c \sim l \cdot p_0^{\beta-1}$, where $l$ is the scattering length along the laser beam.

The experimental arrangement is shown in Figure 1(b). A cw laser beam (532 nm, $\approx$ 90 mW) which was expanded to 25 mm diameter was focused into the gas jet by a lens with a focal length of 30 cm. The Rayleigh signal scattered from clusters was detected in 90° arrangement. The scattered light was imaged onto a photomultiplier tube module (Hamamatsu H6779 series) with a bandpass filter, which allows time-resolved detection of spatially integrated cluster distribution. The output of photomultiplier was recorded by a digital oscilloscope (Tektronix DPO4104B-L). The valve was opened for at least 3 ms duration, which is needed for obtaining steady state gas flow. The measured scaling laws did not change significantly with the laser axis-nozzle distance $x$, therefore we chose the distance of x = 1.5 mm which distance has been characterized in [9, 12, 15].

The monitoring of spatial distribution of clustering was obtained by a CCD camera (PCO Pixelfly) with an exposition time of 1 ms (with a spatial resolution of $\approx$ 30 µm/pixel determined by the pixel size, as assured by the F/12 focusing of the $TEM_{00}$ beam). In this case the nozzle position was changed with a translation stage perpendicular to the laser axis at a distance of 0-4 mm. The spatial distribution and propagation of clusters were investigated in 12 bar Ar and 2 bar Xe. The choice of these values of pressure were reasonable for the two gases, because in such ways the CCD camera didn't saturate at any distances in the range of 0-4 mm, even though in Xe the signal is far higher than in Ar.

Depending on the gas, backing pressure and nozzle the interval of $\Gamma^* \approx 10^3 - 1.5 \times 10^5$ was scanned which is theoretically the range for the original- and corrected Hagena's law [8, 9].

## 4. Results

The sections of CCD images from Rayleigh signals are put on top of each other changing the laser axis-nozzle distance $x$ with a step of 0.1 mm. Figure 2 shows thus the spatial propagation of the clusters for each nozzle both in Ar and Xe. Note that the Rayleigh signal could hardly be seen in Ar for the de Laval micronozzle. It can be immediately seen that whereas the micronozzle and the supplemental converging nozzle give a nice flat-top distribution, there is some modulation in Ar for the cylindrical nozzle and annular distribution can be observed for the de Laval nozzle of 1 mm.

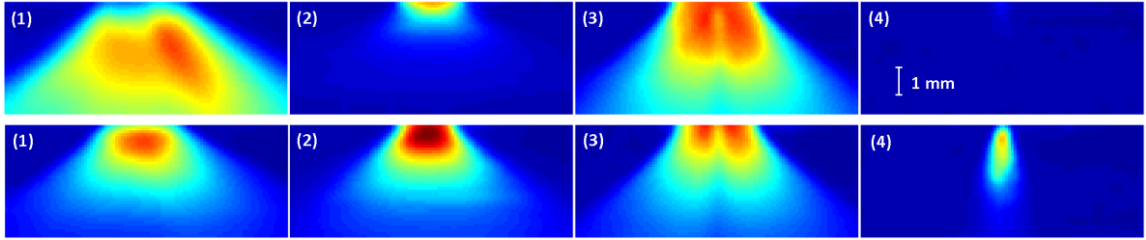

**FIG. 2.** Rayleigh signal for argon cluster jets for nozzles (1), (2), (3) and (4) with 12 bar backing pressure (top). Xenon cluster jets with 2 bar backing pressure (bottom).

The total Rayleigh signal integrated along the z axis shows that some distance from the nozzle exit is necessary for clustering which is most obvious in case (1) when the expansion of the gas starts practically at the orifice. The integrated signal and thus the cluster size depends on this distance ($x$). A plateau can be observed after ~1 mm for this cylindrical nozzle. Case (2) shows a similar behaviour in xenon, i.e. for the case with larger clusters. Quantitative evaluation can be obtained from the measurements by photomultiplier detector. The pressure dependences of cluster sizes from the measured time-integrated signal are shown between 2 and 20 bar for Ar and Xe in the case of the four nozzles in Figure 3(a) and (b). Every data point was averaged at least for 10 measurements. The smallest detectable signal was measured in 1 bar Ar with the de Laval nozzle of 1 mm, supposing that the cluster size is 100 in this case. It can be immediately seen that the power dependence is different for the different nozzles, which is demonstrated by the crossing of different curves for Xe.

The measured data were fitted by power functions and the obtained $\beta - 1$ power is given in Table 1 for the different gases and nozzles. The power dependence of the cluster size is then given by $\beta - 1$ which can be compared with the analytical scalings. It is remarkable that the argon data gives for the cylindrical and the supplemental nozzle higher powers than the Hagena scaling. In case of Xe the supplemental converging nozzle gives higher values for $\beta - 1$, thus a pressure dependence which is much steeper than expected from the Hagena scaling. Interesting is the case of de Laval nozzle of 1 mm for which the simple power law fitting is not optimal, as one can distinguish the range of 2-12 bar and that of 12-20 bar.

| Nozzle: | Cylindrical | Supplemental Converging | De Laval 1 mm | De Laval 0.15 mm |
|---|---|---|---|---|
| $\beta - 1$ (Ar) | $2.65 \pm 0.09$ | $3.46 \pm 0.29$ | $2.18 \pm 0.05$ | - |
| $\beta - 1$ (Xe) | $1.72 \pm 0.04$ | $3.68 \pm 0.34$ | $1.7 \pm 0.23$ | $1.92 \pm 0.13$ |

**Table 1.** The $\beta - 1$ power derived from our Rayleigh signals for the four nozzles in argon and xenon which can be compared with the $\beta - 1 \approx 1.8$ scaling [9].

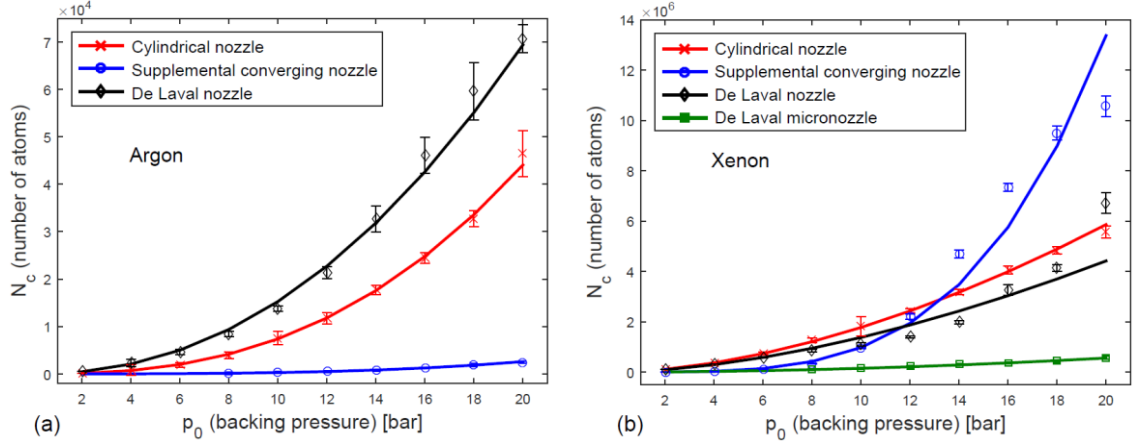

**FIG. 3.** Backing pressure dependences of the cluster size $N_c$ for different nozzles in argon (a) and in xenon (b). Solid lines represent the fitted power functions.

## 5. Discussion

Most of the experiments with gas jets require a homogeneous, possibly flat-top density distribution. As clustering is a relatively slow process, some distance is needed for the condensation, that is why for the cylindrical nozzle - in which case the expansion starts only at the orifice - the cluster size reaches the maximum value farther from the exit. Density modulations were foreseen in the jet centre for certain nozzle contours according to simulations [14] in monomer gases which are clearly observed in Figure 2. It was shown in previous experiments with Ar [9, 16] that cluster formations may result in annular density distribution for supersonic de Laval nozzles. In our case there was no density drop in the jet center for the nozzle of exit diameter 0.15 mm, whereas for the exit diameter 1 mm a hollow distribution can be seen, thus in order to carry out experiments with a flat-top profile it is necessary a precise mounting of the laser-gas jet position along the y-axis.

Based on Table 1 we can compare the experimental results with the scalings. Except for the supplemental converging nozzle the Xe data are in agreement with the scaling of Dorchies et al. [9]. On the other hand larger exponents were found for Ar, i.e. the original Hagena scaling is a better approximation. Different is the case of the supplemental converging nozzle, for which a larger exponent was observed in both gases corresponding to a stronger increase of cluster size with pressure. The probable reason of it is that the process of clustering proceeds in the 12 mm long cylindrical volume, which concentrates the gas density given by the commercial nozzle. In order to obtain large cluster sizes, the supplemental nozzle is preferred with xenon gas. The case of the 1 mm de Laval nozzle is interesting because in Xe a single power scaling is hardly adapted: A lower $N_c \sim p_0^{1.46}$ dependence is more probable for pressures below 12 bar and a steep scaling with $N_c \sim p_0^{3.1}$ can be better fitted between 12 and 20 bar.

Interesting results can be obtained from the comparison of data in Ar and in Xe. Note that the most previous experiments used preferentially Ar clusters [9, 10]. According to the Hagena's prediction the ratio of cluster size in Ar and Xe depends only on one material parameter, i.e. the condensation parameter ($k$) of the gas. Therefore the ratio of the number of atoms should be independent from the backing pressure. This is in contrast of the results shown in Figure 4. It can be clearly seen, that the measured ratio is not constant, it varies strongly. It is practically always higher - especially for the supplemental converging nozzle - than obtained from Hagena's law $N_{c,\text{Xe}}/N_{c,\text{Ar}} = (k_{\text{Xe}}/k_{\text{Ar}})^{2.35} \approx 40$. According to these measured ratios the material dependences cannot be described by a single parameter ($k$) or this parameter can depend on the nozzle geometry.

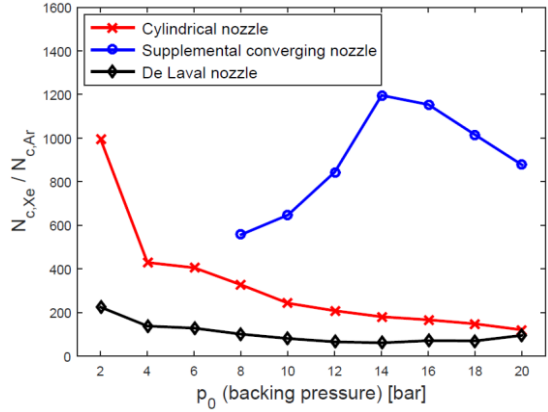

**FIG. 4.** Cluster size ratio for xenon and argon depending on the backing pressure in the case of three nozzles. The data for the supplemental converging nozzle is divided by 5 for a better comparison.

In our previous HHG experiments [12] we estimated the cluster density from the measured spectral redshift of the 21$^{st}$ harmonic based on a nanoplasma model. Those measurements were carried out using a shorter opening of the valve, in which case the Rayleigh signal and the corresponding atom number of a cluster was lower. The number of atoms for 12 bar backing pressure was $\approx 2 \times 10^4$ as compared with the $\approx 2 \times 10^6$ herewith. This corresponds to a cluster radius of $\approx 8$ nm with the assumed atomic distance in the cluster. In the nanoplasma model the spectral shift is the function of a product of $N_c$ with the cluster density, resulting in $\approx 6 \times 10^{15}$ clusters/cm$^3$ under the experimental conditions in [12]. Thus Rayleigh scattering allows a more accurate, separate determination of the cluster radius.

## 6. Conclusions

Summarizing the results we could confirm that Rayleigh scattering is an essential diagnostics for characterizing the size of clusters. It was demonstrated that the cluster sizes depend strongly on the actual geometry of the nozzles and different shaped nozzles provide different scaling for the pressure dependence of the cluster size.

It was shown that the Hagena formula and its corrected version are not sufficient for the characterization of the clusters, especially for the material dependence. Different dependence of the cluster size ratios for Xe and Ar on the pressure for different nozzles shows that the deviation cannot be simply described with one condensation parameter but geometrical effects should be added as well.

The cluster sizes obtained from Rayleigh scattering diagnostics improve the estimations of the nanoplasma model. Measuring simultaneously the cluster size with the redshift of high harmonics allows one to determine the density of the clusters in the interaction volume.

### Acknowledgments

This work has been carried out within the framework of the EUROfusion Consortium and has received funding from the Euratom research and training programme 2014-2018 under Grant Agreement No. 633053. The views and opinions expressed herein do not necessarily reflect those of the European Commission. The work was also supported by the Munich Center for Advanced Photonics (MAP), DFG Project Transregio TR18.